\documentclass[aip,jap,amssymb, amsmath, numerical,reprint]{revtex4-1}
\usepackage{graphicx}
\usepackage{epstopdf}
\usepackage{dcolumn}%
\usepackage{bm}%
\newcommand{\rfig}[1]{Fig.~\ref{#1}}
\newcommand{\Vt}{$V_{\text{tune}}~$}
\newcommand{\unit}[1]{\,\text{#1}}
\usepackage[]{natbib} %
\usepackage[colorlinks=true,linkcolor=blue]{hyperref}%
\expandafter\ifx\csname package@font\endcsname\relax\else
 \expandafter\expandafter
 \expandafter\usepackage
 \expandafter\expandafter
 \expandafter{\csname package@font\endcsname}%
\fi
\begin{document}

\title{30\,GHz-voltage controlled oscillator operating at 4\,K\\}%

\author{Arne Hollmann}%
\email[email to: ]{arne.hollmann@rwth-aachen.de}
\affiliation{JARA-FIT Institute for Quantum Information, Forschungszentrum J\"ulich GmbH and RWTH Aachen University, D 52074 Aachen, Germany}
\author{Daniel Jirovec}
\affiliation{Institute of Science and Technology Austria, Am Campus 1, 34
00 Klosterneuburg, Austria}
\author{Maciej Kucharski}
\affiliation{IHP, Im Technologiepark 25, 15236 Frankfurt (Oder), Germany}
\author{Dietmar Kissinger}
\affiliation{IHP, Im Technologiepark 25, 15236 Frankfurt (Oder), Germany}
\author{Gunter Fischer}
\affiliation{IHP, Im Technologiepark 25, 15236 Frankfurt (Oder), Germany}
\author{Lars R. Schreiber}
\affiliation{JARA-FIT Institute for Quantum Information, Forschungszentrum J\"ulich GmbH and RWTH Aachen University, D 52074 Aachen, Germany}
\date{\today}%
\begin{abstract}
 Solid-state qubit manipulation and read-out fidelities are reaching fault-tolerance, but quantum error correction requires millions of physical qubits and thus a scalable quantum computer architecture. To solve signal-line bandwidth and fan-out problems, microwave sources required for qubit manipulation might be embedded close to the qubit chip, typically operating at temperatures below 4\,K. Here, we perform the first low temperature measurements of a 130\,nm BiCMOS based SiGe voltage controlled oscillator. The device maintains its  functionality from 300\,K to 4\,K. We determined the dependence of frequency and output power on temperature and magnetic field up to 5\,T and measured the temperature influence on noise performance. While the output power tends to increase, the frequency shift is 3\,\% for temperature and 0.02\,\% for the field dependence, respectively, both relevant for highly coherent spin qubit applications. We observe no improvement on output noise, but increased output flickering.
\end{abstract}
\maketitle
\tableofcontents

\section{Introduction}
The performance of individual solid-state qubits are reaching a level that is sufficient to perform quantum error correction \cite{Veldhorst2014, Watson2017, Muhonen2015, Yoneda2017,  Barends2014}. Until today, all concepts of solid-state quantum processors share the need of cryogenic temperatures below 1\,K. 
However, any quantum processor has to be interfaced by a classical control hardware to manipulate and read out the qubits. It has to be fast compared to the time scale of decoherence, low-noise and scalable in terms of required lines, size and power consumption \cite{VanDijk2018}. Fulfilling these requirements for an enormous number of signal lines is a scale-up challenge. Besides multiplexing \cite{Puddy2015, Hornibrook2015, Al-Taie2015, Veldhorst2017} concepts, it might be unavoidable to place part of control hardware close to the qubits and thus at cryogenic temperature. Power-consuming control hardware might be placed at temperatures between 1~K and 4~K, since the available cooling power is unproblematic in this range. The qubit chips either have to be thermally isolated e.g. by superconducting signal lines or have to be operated at elevated temperature \cite{Vandersypen2016}. 
\\ Within the last years there have been several approaches to start developing a cryogenic control hardware. The use of cryogenic digital-to-analogue converters  \cite{Rahman2014}, analogue-to-digital converters \cite{Okcan2010} and field-programmable gate arrays \cite{Homulle2017, ConwayLamb2016} enables the possibility to process data in close vicinity of the qubit and avoid a high rate data transfer between quantum processor and room temperature control electronics. \\ Depending on the specific qubit realisation, control and read-out schemes require high frequency signals in the 100\,MHz range \cite{Cerfontaine2014} up to tens of Gigahertz \cite{Schreiber2011, Barends2014,  Zajac2017, Watson2017, Yoneda2014, Yoneda2017}. It might also become beneficial to develop cryogenic high frequency sources which need to be low-noise and tunable in frequency space, in order to be able to address different qubits and to enable high-fidelity qubit operation\cite{VanDijk2018}. \\
 Here, we test the performance of a commercial heterojunction bipolar transistor (HBT) based voltage controlled oscillator (VCO) at 4\,K. SiGe HBTs are known to work at cryogenic temperatures thanks to their high doping concentration \cite{Cressler2005, Cressler2010, Chakraborty2014} and thus could be a candidate for future qubit control electronics. The device did show its full functionality at 4\,K. We measured that the oscillator's phase noise is mainly dominated by input noise coupling to the system. The carrier frequency is slightly affected by the external magnetic field. 
\\ The paper is structured as follows: In Sec. \ref{sec:device} we shortly present the device under test and the measurement set-up. In Sec. \ref{sec:measurements} the low temperature characterisation is presented including phase noise measurements and magnetic field dependence of the output signal.

\section{Device description}
\label{sec:device}
\begin{figure}
\centering
\includegraphics[]{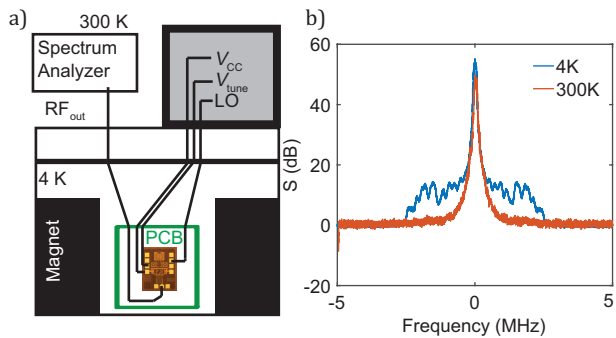}
\caption{a) Schematic of the measurement set-up. The VCO is bonded to a PCB and can be measured at 4.2\,K and in a magnetic field up to 6\,T. Supply voltages are applied with a home built low noise DC source. b) Average power spectrum of 20 measurements taken at room temperature (red) and at 4\,K (blue) as a function of frequency offset from the carrier frequency. The peak position is shifted from 29.81\,GHz at 300\,K to 30.76\,GHz at 4\,K.}
\label{setup}
\end{figure}

In this work we tested a wide band, low phase noise LC-voltage controlled oscillator (VCO) at 4\,K. The device was fabricated in a SiGe 130\,nm BiCMOS technology of Innovations for High Performance Microelectronics (IHP) \cite{Kucharski2016}. The VCO can be tuned continuously from 29.6\,GHz to 32.4\,GHz (at room temperature) with multiple varactor banks. Additionally, by using a switch, an upper frequency band from 32.0\,GHz to 35.5\,GHz can be used. This results in a large total tuning range of 18.1\%. The change between the two frequency bands is realised by switch coupled inductors. The switch consists of heterojunction bipolar transistors (HBTs) allowing to switch between two inductances $L_1$ and $L_2$.  A schematic of the circuit is shown in ref. \cite{Kucharski2016}. The circuit occupies an area of 0.1\,$\text{mm}^2$. For operation, a supply voltage of $V_{\text{CC}} =3\,\text{V} $ is needed. 
\\ 
The VCO is bonded to a RF-PCB that enables feeding the VCO with the corresponding voltages as well as tapping the RF-signal. The PCB with the VCO is tested in a 4\,K set-up consisting of a dipstick and a helium vessel with a 6\,T magnet. The magnetic field can be applied perpendicular to the device under test. The output signal of the VCO is measured with a 40\,GHz spectrum analyser (Fig.~\ref{setup} (a)). 

\section{Electrical characterisation}
\label{sec:measurements}
\begin{figure}[t]
\centering
\includegraphics[]{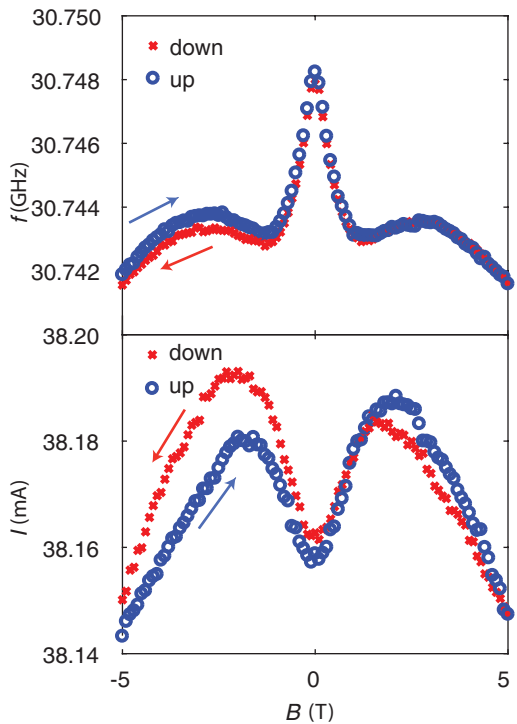}
\caption{Top: Carrier frequency as a function of applied magnetic field. The magnetic field is firstly swept upwards from -5\,T to 5\,T and than in reverse direction from 5\,T to -5\,T. Bottom: Current consumption for the same measurement. }
\label{lastfig}
\end{figure}

Fig.~\ref{setup} (b) shows a spectrum averaged over 20 single traces taken at room temperature (red curve) and at 4\,K (blue curve). The power has been normalised to the noise level. At 300\,K we measure an output power of $-31.5$\,dBm. The spectrum has  a sharp peak with a full width half maximum of roughly 200\,kHz and centre frequency of 29.81\,GHz. 
In comparison, at 4\,K the carrier frequency is shifted by almost 1\,GHz to 30.76\,GHz. The peak width is similar to the room temperature measurement, however, at low temperature, next to the main peak we measure small oscillations with an abrupt edge at $\pm 2.5\,\text{MHz}$ distance from the centre peak. The power of the oscillations is 45\,dB smaller than the maximum of the signal. This effect could be attributed to small, slow output fluctuations on the sinusoidal output signal. An orders of magnitude smaller effect is also present at room temperature but not visible in Fig. \ref{setup} (b). By assuming a square-like flickering modulation on the output signal, we estimated that the amplitude of this effect is a factor of 100 smaller than the amplitude of the output signal. Generally, the behaviour of the VCO should be improved when operated non freely but in a closed loop. 
\begin{figure}[tpb]
\centering

\includegraphics[]{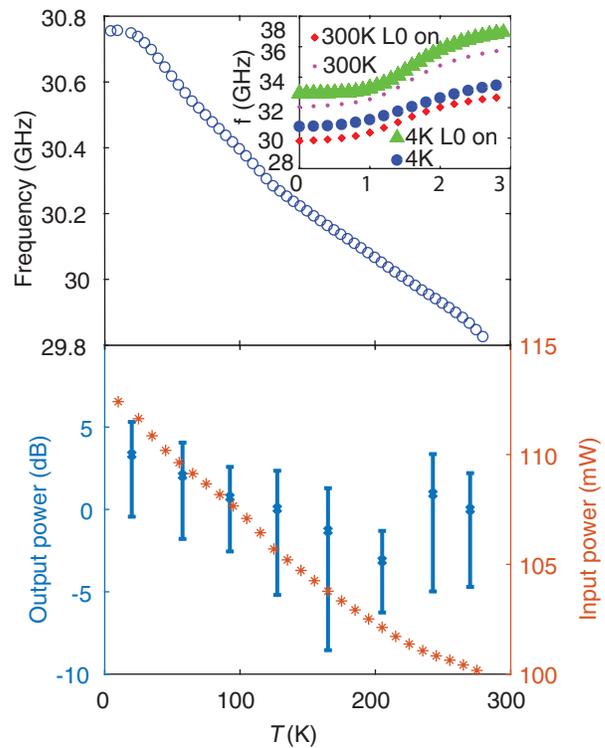}
\caption{ Top: Carrier frequency versus temperature, measured during cool down. Inset: Carrier frequency as a function of $V_{\text{tune}}$ at 300\,K (red), 300\,K L0 on, 4\,K (blue) and 4\,K with L0 on (green). Bottom: Output (blue) and input (orange) power of the VCO as a function of temperature. The output power values are averaged over six measurements, the error bar indicates one standard deviation.}
\label{funct}
\end{figure}
\\To further investigate the effect of temperature, we measured the power spectrum during cool down of the device. The carrier frequency shifts almost linearly between 300\,K and 20\,K. Below 20\,K the carrier frequency remains constant at around 30.76\,GHz (Fig.~\ref{funct}). Both at room temperature and at 4\,K, the frequency can be varied by almost 3\,GHz when applying a tuning voltage $V_{\text{tune}}$ between 0\,V and 3\,V to the varactor bank (inset Fig.~\ref{funct} top). An additional frequency shift of 3\,GHz can be induced by applying a voltage to the switch-coupled inductors. 
In total, a frequency window of almost 6\,GHz is accessible. We conclude that both  the varactor and the coupled inductor operate down to 4\,K. Beside a change in frequency, we measured an increase in the power consumption of 28\% when reducing the temperature to 4\,K (Fig.~\ref{funct}). The power of the output signal is increased by roughly 4\,dB. \\
The oscillator's phase noise is another important  figure of merit, as a clean signal with low phase or frequency noise is essential for any application.
\begin{figure}[tpb]
\centering
\includegraphics[]{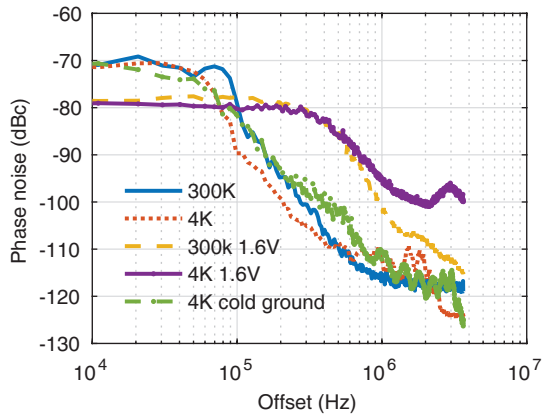}
\caption{Phase noise versus offset from the carrier frequency at 300\,K and 4\,K where the $V_{\text{tune}}$-pin was set to lab ground (blue and red respectively) and where it was bonded to cold ground (green). No large variation can be detected, implying that the VCO is limited by thermal noise. Setting $V_{\text{tune}}$ to 1.6\,V shows an increased phase noise (yellow and purple).}
\label{phase noise}
\end{figure}
\\In semiconducting oscillators one typically distinguishes device noise and modulation noise. Device noise is contributed by the device itself consisting of $1/f$-noise only being relevant at low frequencies as well as shot noise and thermal noise which are dominant for high frequency application. Additionally, the output signal is affected by noise of several DC voltages such as the supply voltage $V_{\text{CC}}$, the tuning voltage $V_{\text{tune}}$ and the substrate voltage \cite{Herzel1999, Herzel2004} being referred as modulation noise.  Here, we measured phase noise of the output signal as a function of temperature, which should ideally be decreased for low temperature values. 
\\The phase noise $S_{\omega}$ can be extracted easily from the power spectrum of the output signal. In the following, we will refer to the single-side band power $S_\text{SSB}$ in one Hertz bandwidth, normed to the power of the peak as phase noise:
\begin{equation}
\mathcal{L} (\omega)[\text{dBc/Hz}]= \frac{S_{\text{SSB}}(\omega, \Delta f= 1\,\text{Hz})}{\int_{peak} \text{d} \omega \;S_{\text{SSB}}}.
\label{eq:Lofomega}
\end{equation}
\\The resulting phase noise as a function of frequency offset from the carrier frequency is depicted in Fig.~\ref{phase noise}. The blue solid line reveals the phase noise at room temperature dropping from -80\,dBc at 100\,kHz to roughly -115\,dBc at 1\,MHz. The overall phase noise is small and comparable to other low noise VCOs \cite{Chien2007, Chen2005,Chen2013}. We now compare the result to a similar measurement taken at 4\,K (red dotted line). At 100\,kHz the phase noise is lower (-90\,dBc), at 1\,MHz slightly higher (-110\,dBc) with respect to the room temperature measurement. For frequencies in the MHz-range, the oscillations also noticed in \rfig{setup} become visible. 
\begin{figure}[tpb]
\centering
\includegraphics[]{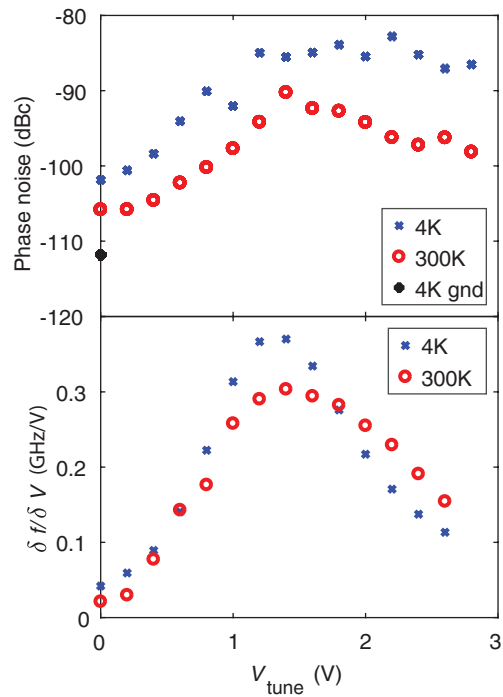}
\caption{Top: Phase noise at a constant offset of 1\,MHz versus \Vt at 300\,K and 4\,K. The phase noise is strongly increased for increasing \Vt~up to 1.4\,V,  being reduced for $V_{\text{tune}}\geq1.4\,\text{V}$ at 300\,K and staying almost constant at 4\,K, respectively. Bottom: For comparison, the derivative of the frequency tuneability \rfig{funct} (b) inset with respect to \Vt~is displayed for the two operation temperatures.  The sensitivity of the carrier frequency with respect to \Vt~matches the phase noise amplitude plotted in the upper panel.}
\label{phase vtune}
\end{figure}
\\ The behaviour of the phase noise remains similar also when connecting the $V_{\text{tune}}$-port to the grounded cold finger of the fridge (referred to as cold ground, Fig. \ref{phase noise}) instead of the lab ground at the dipstick (red trace in Fig. \ref{phase noise}). Vice versa, the phase noise increases consistently, both at room temperature and 4\,K, when applying a finite voltage to the $V_{\text{tune}}$-port ($V_{\text{tune}}=1.6\,\text{V}$ in \rfig{phase noise}). These results strongly suggest that the phase noise of the VCO is dominated not by the characteristics of the device itself, but by the input noise due to $V_{\text{tune}}$ (modulation noise)\cite{Herzel2004}. 
\\To further analyse the sensitivity of phase noise to $V_{\text{tune}}$ we measured the phase noise for different tuning voltages. A low-noise, home built DC source was connected to the $V_{\text{tune}}$-pin and the voltage was varied over the full range from 0\,V to 3\,V. The resulting phase noise at a constant offset of 1\,MHz from the carrier frequency is depicted in Fig~\ref{phase vtune}. Increasing the tuning voltage up to 1.4\,V leads to an increase in noise of 15\,dB. For $V_{\text{tune}} \geq 1.4\,\text{V}$ the noise decreases again at 300\,K, however at 4\,K it stays almost constant. Even applying zero voltage increases the noise by 10\,dB compared to $V_{\text{tune}}$ set to ground. Furthermore, we compared these measurements with the derivative of the carrier frequency of the output signal with respect to \Vt from the measurement data in \rfig{funct} (b) (inset). As depicted in \rfig{phase vtune} the qualitative behaviour of the curves is similar. 
Within the region where the carrier frequency is  most sensitive to \Vt the VCO is also most sensitive to noise of the DC source. 
\\ The natural qubit environment does not only require low temperature but usually also the presence of a magnetic field. Especially, for encoding spin qubits magnetic fields in the range of 1\,T are needed \cite{Loss1998, Kawakami2014, Vandersypen2016}. Thus, the effect of magnetic field on the performance of electronics in close proximity to the qubit-chip has to be considered.
In  the following, we study the effect of magnetic field on the carrier frequency of the VCO. The magnetic field has been applied perpendicular to the VCO and swept from $-5\,\text{T}$ to $5\,\text{T}$ and finally reversely, from $5\,\text{T}$ to $-5\,\text{T}$. At $5\,\text{T}$ and $-5\,\text{T}$ the carrier frequency is reduced by 6\,MHz compared to zero magnetic field (a factor of ~$10^{-4}$) (\rfig{lastfig}). The $B$-field dependence of the carrier frequency reveals a saddle point at around $\pm 2\unit{T}$. Furthermore, the up-sweep and down-sweep do not coincide exactly. A tiny hysteresis is  visible on the order of $10^{-5}$.  The current consumption reveals two side peaks at $\pm 2\unit{V}$ (\rfig{lastfig}), but here again the overall change is small (in the ppm range). The magnetic field dependence on the frequency is a tiny fraction of the output frequency. This is not a severe limitation, since usually the magnetic field is fixed in spin qubit experiments but has to be taken into account for high fidelity spin qubits in e.g. $^{28}$Si \cite{Yoneda2017}.

\section{Conclusion}
In this work we presented the first low-temperature measurements of a high frequency (30\,GHz) VCO. The device showed complete functionality, even at 4\,K. By varying the control voltage \Vt and using the base band switch a frequency band of 6\,GHz is accessible. Compared to room temperature the carrier frequency is shifted by almost 1\,GHz when operating the device at cryogenic temperatures. The phase noise of the oscillator has been analysed at room temperature and at 4\,K respectively. Since the noise is dominated by modulation noise, e.g. of the tuning voltage, no significant effect of the temperature could be measured. \\However, a power consumption in the order of 100\,mW restricts the use of this device to the 4\,K stage of a dilution refrigerator. Nonetheless, the SiGe-HBT BiCMOS technology can be a promising platform to start the development of a cryogenic classical control technology for a scalable quantum computer.
%\addcontentsline{toc}{section}{Acknowledgement} 

\section*{Acknowledgement}
We would like to thank H. Bluhm for useful discussions and F. Haupt for proof reading this manuscript.

\end{document}